\ifx\pdfoutput\undefined\else\pdfoutput=1\fi

\documentclass[aps,prl,showpacs,twocolumn,superscriptaddress,longbibliography]{revtex4-1}

\usepackage{amsmath,amssymb,bm}
\usepackage{graphicx}                 
\DeclareGraphicsExtensions{.pdf,.png,.jpg}
\usepackage{xcolor}

\usepackage{CJKutf8}

\usepackage[colorlinks=true,citecolor=blue,linkcolor=blue,urlcolor=blue]{hyperref}

\begin{document}
\begin{CJK*}{UTF8}{gbsn}

\preprint{APS/123-QED}

\title{Experimental demonstration of genuine quantum information transmission through completely depolarizing channels in a superposition of cyclic orders}

\author{Yaxin Wang}
\affiliation{State Key Laboratory of Optoelectronic Materials and Technologies and School of Physics, Sun Yat-sen University, Guangzhou 510000, China}


\author{Linxiang Zhou}
\affiliation{State Key Laboratory of Optoelectronic Materials and Technologies and School of Physics, Sun Yat-sen University, Guangzhou 510000, China}

\author{Tianfeng Feng}
\email{fengtf@hku.hk}
\affiliation{QICI Quantum Information and Computation Initiative, School of Computing and Data Science,
 The University of Hong Kong, Pokfulam Road, Hong Kong}

\author{Hanlin Nie}
\affiliation{Centre for Quantum Technologies, National University of Singapore, Block S15, 3 Science Drive 2, 117543, Singapore}

\author{Ying Xia}
\affiliation{State Key Laboratory of Optoelectronic Materials and Technologies and School of Physics, Sun Yat-sen University, Guangzhou 510000, China}

\author{Tianqi Xiao}
\affiliation{State Key Laboratory of Optoelectronic Materials and Technologies and School of Physics, Sun Yat-sen University, Guangzhou 510000, China}

\author{Juntao Li}
\affiliation{State Key Laboratory of Optoelectronic Materials and Technologies and School of Physics, Sun Yat-sen University, Guangzhou 510000, China}

\author{Vlatko Vedral}
\affiliation{Clarendon Laboratory, University of Oxford, Parks Road, Oxford OX1 3PU, United Kingdom}

\author{Xiaoqi Zhou}
\email{zhouxq8@mail.sysu.edu.cn}
\affiliation{State Key Laboratory of Optoelectronic Materials and Technologies and School of Physics, Sun Yat-sen University, Guangzhou 510000, China}
\affiliation{Hefei National Laboratory, Hefei 230088, China}

\date{\today}

	\begin{abstract}
	 A major challenge in quantum communication is addressing the negative effects of noise on channel capacity, especially for completely depolarizing channels, where information transmission is inherently impossible. The concept of indefinite causal order provides a promising solution by allowing control over the sequence in which channels are applied. We experimentally demonstrate the activation of quantum communication through completely depolarizing channels using a programmable silicon photonic quantum chip. By implementing configurations based on {superposition of cyclic orders, a form of indefinite causal order,} we report the first experimental realization of genuine quantum information transmission across multiple concatenated completely depolarizing channels. Our results show that when four completely depolarizing channels are combined using {the superposition of cyclic orders}
     , the fidelity of the output state is $0.712 \pm 0.013$, significantly exceeding the classical threshold of $2/3$. Our work establishes indefinite causal order as a powerful tool for overcoming noise-induced limitations in quantum communication, demonstrating its potential in high-noise environments and opening new possibilities for building robust quantum networks.
	\end{abstract}
	\maketitle

\section*{Introduction}
Reliable information transmission is a fundamental goal of communication, yet noise is an inherent limitation. Common types of noise include thermal noise \cite{CallenWelton1951}, quantum decoherence noise \cite{Zurek2003RMP,BreuerPetruccione2002}, and noise from completely depolarizing channels (CDCs) \cite{King2003Depol,NielsenChuang2010,Shannon1948}. These noise sources severely limit channel capacity and, in extreme cases, may result in complete information loss. Thus, overcoming noise-induced limitations in information transmission remains a key challenge in quantum communication.

Indefinite Causal Order (ICO) is a novel quantum resource that harnesses the quantum superposition of causal orders, fundamentally overcoming the limitations of a fixed causal order \cite{Chiribella2013QCS, Ebler2018,salek2018quantum,Chiribella2021,Nie2022ArXiv,Procopio2020,Procopio2015,Araujo2014, Rubino2017,Goswami2018,Procopio2019,Renner2022a,Sazim2021Classical,Guo2020,Goswami2020,Rubino2021,Guerin2016,Wei2019,Zhao2020,Yin2023,Nie2022,Zhu2023,Wu2025,Chiribella2018, Chiribella2022, Taddei2021,Dieguez2023,Bavaresco2021, Quintino2019,Felce2020,Zhao2020PRL,Yin2023NatPhys,Liu2023PRL,Das2022PRSA}. This mechanism enhances quantum channel capacity and permits information transfer through otherwise non-transmissive channels, such as completely depolarizing channels, thereby enabling communication where it was previously impossible \cite{Ebler2018,Chiribella2021,Guo2020,Goswami2020}.

Recent experimental studies on discrete optical platforms provide evidence that ICO can activate information transmission in CDCs. Experimental results indicate that for $N=2$, ICO enables classical information transmission via CDCs, yet the fidelity remains insufficient for quantum information transmission \cite{Guo2020,Goswami2020}. Theoretical studies predict that when $N=4$, the fidelity of the transmitted quantum state surpasses the classical quantum teleportation limit of $F_{\mathrm{cl}}=2/3$ \cite{Bruss1999,Hayashi2005,Hu2023,Ren2017}, thus making genuine quantum information transmission possible \cite{Chiribella2021}. However, the experimental realization of an $N=4$ channel configuration introduces significant technical challenges, including system stability, phase control, and noise accumulation, that surpass the capabilities of traditional discrete optical platforms.

In this paper, we designed and fabricated a programmable silicon photonic chip, which serves as the platform for the first experimental demonstration of quantum information transmission via ICO in CDCs. We investigated quantum state transmission under ICO across cyclic order channel configurations \cite{Chiribella2021,Nie2022ArXiv}. The results show that when a quantum state propagates through four CDCs under ICO, the output fidelity reaches $0.712 \pm 0.013$, surpassing the classical threshold of $2/3$, thereby enabling quantum information transmission. These results establish ICO as a viable mechanism for quantum information transmission through CDCs, highlighting its potential for quantum communication in high-noise environments.

\section*{Results}

\textbf{Setup.---}
In quantum communication, the CDC represents an extreme case of quantum channel noise. It is mathematically defined as \cite{Ebler2018,Chiribella2021}
\begin{align}
\mathcal{D}(\rho) = \frac{1}{d^{2}} \sum_{i=0}^{d^{2}-1} U_{i} \rho U_{i}^{\dagger} = \frac{I}{d},
\end{align}
where ${U_i}$ denotes a set of orthogonal unitary operators, and $I$ is the $d$-dimensional identity matrix. This expression shows that regardless of the input state $\rho$, the CDC always outputs the maximally mixed state $\frac{I}{d}$, resulting in a complete loss of quantum information. Typically, a quantum state remains maximally mixed after multiple sequential CDC applications, thus preventing any possibility of information transmission. As illustrated in Fig.1(a), a quantum state $\rho$ sequentially traverses multiple CDCs (labeled $A,B,C,D$). The first CDC application instantly collapses it into the maximally mixed state $\frac{I}{d}$, with subsequent applications having no effect. Independent of the number of channels, a CDC arranged in a fixed causal order cannot facilitate quantum information transmission.

A CDC under a fixed causal order cannot support quantum information transmission. However, if the order of channel operations is placed in a quantum superposition, this limitation can be overcome. This ICO mechanism allows a quantum state to traverse multiple channel orderings simultaneously rather than following a fixed sequence. The resulting quantum interference effects can partially recover the quantum information lost due to CDC operations.

A natural approach to realizing indefinite causal order (ICO) with $N$ quantum channels is to construct a coherent superposition over all N! possible permutations of the channel order\cite{Procopio2019,Renner2022a}. While this fully represents the possible causal structures among the channels, it requires a control system of dimension $N!$, and the associated resource requirements grow rapidly with $N$.To address this challenge, Chiribella et al.~\cite{Chiribella2021} proposed the cyclic order model. Instead of superposing all permutations, this approach considers only the $N$ cyclic permutations of the channels. The control system is thereby reduced to dimension $N$, while the model still captures the essential interference effects of indefinite causal order. It has been shown that this reduced configuration suffices to activate quantum communication through completely depolarizing channels.

A representative implementation of the cyclic order model can be realized using a quantum switch\cite{Chiribella2013QCS,Ebler2018}. As shown in Fig.1(b), the quantum switch correlates the order of channel operations with the state of a control qudit, allowing the input state $\rho$ to pass through the CDC in different sequences. When the control state is $|0\rangle_c$, the channels act in the order $A \rightarrow B \rightarrow C \rightarrow D$; when the control state is $|1\rangle_c$, the order becomes $B \rightarrow C \rightarrow D \rightarrow A$; similarly, for $|2\rangle_c$ and $|3\rangle_c$, the respective orderings are $C \rightarrow D \rightarrow A \rightarrow B$ and $D \rightarrow A \rightarrow B \rightarrow C$. By preparing the control qudit in a superposition state
\begin{align}
|+\rangle_c = \frac{1}{2} (|0\rangle_c + |1\rangle_c + |2\rangle_c + |3\rangle_c),
\end{align}
the input state $\rho$ experiences all four possible channel orderings simultaneously, inducing quantum interference effects. After passing through the channels, the control qudit is projected onto $|+\rangle_c$, ensuring that all channel sequences contribute equally to the final operation. As a result, the overall effect of the channels on the input state $\rho$ is equivalent to an  effective channel, described by \cite{Chiribella2021}
\begin{align}
\mathcal{E}_{0}(\rho) = \frac{N-1}{N-1+d^2} \rho + \frac{d^2}{N-1+d^2} \frac{I}{d},
\end{align}
where $\rho$ is the input state, $\frac{I}{d}$ is the maximally mixed state, $N$ is the number of channels, and $d$ is the system dimension \cite{Chiribella2021}. This equation shows that under ICO, the output state $\rho'$ is a weighted mixture of the input state $\rho$ and the maximally mixed state $\frac{I}{d}$, with the weighting ratio determined by $N$ and $d$.

For a two-dimensional quantum system ($d=2$), the fidelity of the output state increases with $N$. When $N=2$ or $N=3$, the fidelity is 0.6 and 0.667, respectively, meaning that classical information can be transmitted, but quantum information transmission remains impossible. However, when $N=4$, the fidelity rises to 0.714, surpassing the classical limit for quantum teleportation ($2/3$), confirming that the quantum information transmission capability of the CDC is successfully enabled. Note that, within the cyclic-order model the effective qubit channel is depolarizing, so this fidelity should be input independent and the same for any pure input state.

\begin{figure}[tbp]
		\centering\includegraphics[width=7.6cm]{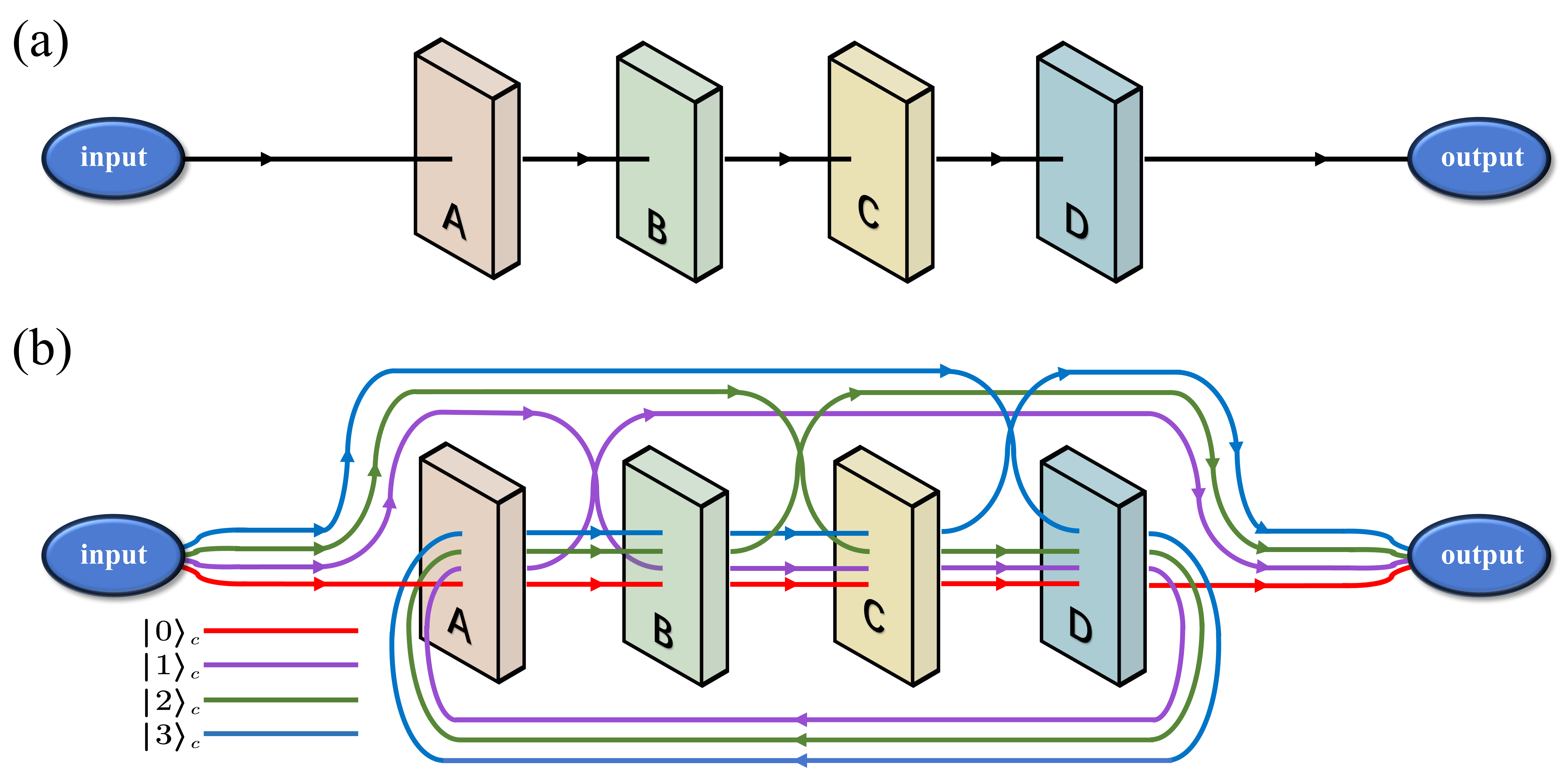}
		\caption{
        The schematic of a quantum state passing through four channels. (a) The quantum state passes sequentially through the channels $A$, $B$, $C$, and $D$ in a fixed order. (b) The quantum state passes through the channels $A$, $B$, $C$, and $D$ in a superposition of four different sequences.
        }
        \label{Fig1a} 
\end{figure}%

    \begin{figure*}[tbp]
	\centering\includegraphics[width=15cm]{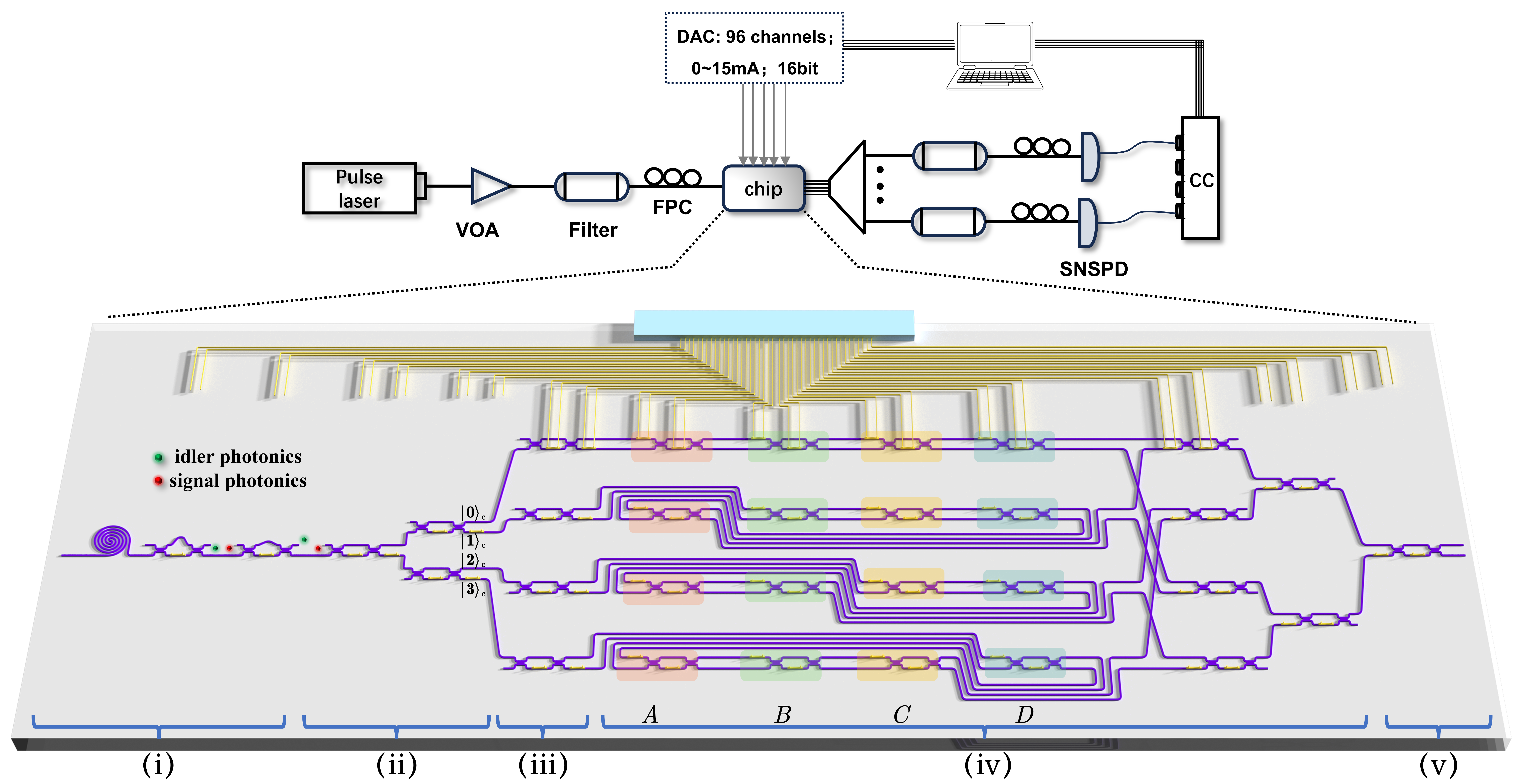}
	\caption{Schematic of the silicon quantum photonic chip and the external setup. The chip comprises five functional regions. (i) generation of single photons. (ii) preparation of the control state $|+\rangle$. (iii) preparation of the input state $\rho$. (iv) the input state $\rho$  passes through four completely depolarizing channels with indefinite causal order. Each channel is configured with equal probability to implement one of the Pauli operations $I$, $X$, $Y$, or $Z$, resulting in 256 different combinations, ranging from $IIII$, $IIIX$, to $ZZZZ$. For each input state $\rho$, these 256 configurations are applied, and the corresponding output data is accumulated, effectively simulating the passage of the quantum state through four completely depolarizing channels. (v) The output state $\rho^{\prime}$ is measured by projecting it onto the six quantum states $0$, $1$, $D$, $A$, $R$, $L$, with each projection being measured for 1 second. The total measurement time for each output state is $1 \times 256 \times 6 = 1536$ seconds. The two-photon coincidence count rate is 116.8/s. VOA, variable optical attenuation; FPC, fiber polarization controller; FA, fiber array; SNSPD, superconducting nanowire single-photon detector; CC, coincidence counting module.}
    \label{Fig2a}
	\end{figure*}

\textbf{Experimental demonstration.---}
To experimentally demonstrate quantum information transmission through CDCs, we developed a programmable quantum optical platform based on a silicon photonic chip. As shown in Fig. 2, this platform integrates all essential components from single-photon generation to quantum state measurement, comprising five functional modules. (i) \textbf{Single-photon generation}. A pulsed laser (1550.12 nm, 80 MHz) is injected into a silicon waveguide spiral, generating entangled photon pairs via four-wave mixing. The pump light is filtered out by an asymmetric Mach-Zehnder interferometer (AMZI), and another AMZI separates the signal (red) and idler (green) photons. The idler photon serves as a trigger, while the signal photon is directed into subsequent circuits. (ii) \textbf{Control state preparation}. The signal photon propagates through a Mach-Zehnder interferometer (MZI) network. By tuning the thermal phase shifters, the control state is prepared as $|+\rangle_c = \frac{1}{2} (|0\rangle_c + |1\rangle_c + |2\rangle_c + |3\rangle_c)$,where $|0\rangle_c$, $|1\rangle_c$, $|2\rangle_c$, and $|3\rangle_c$ correspond to four vertically arranged waveguides on the chip.(iii) \textbf{Input state preparation}. Each of the four waveguides is connected to an MZI circuit with two thermal phase shifters, which are tuned to prepare the photon in a path-encoded single-qubit state $\rho$.
(iv) \textbf{Transmission under indefinite causal order}. The four CDCs ($A, B, C, D$) are implemented using MZIs with phase shifters. By adjusting the phases, the photon undergoes $I$, $X$, $Y$, and $Z$ operations with equal probability, simulating a completely depolarizing channel. The quantum switch enables the input state $\rho$ to traverse the CDCs in four different causal orders, dictated by $|+\rangle_c$
\begin{itemize}
    \item $|0\rangle_c$: $A \rightarrow B \rightarrow C \rightarrow D$
    \item $|1\rangle_c$: $B \rightarrow C \rightarrow D \rightarrow A$
    \item $|2\rangle_c$: $C \rightarrow D \rightarrow A \rightarrow B$
    \item $|3\rangle_c$: $D \rightarrow A \rightarrow B \rightarrow C$
\end{itemize}
Afterward, the control state is projected onto $|+\rangle_c$, ensuring an equal-weight superposition of all causal orders. This results in the final path-encoded output state $\rho'$. (v) \textbf{Output state measurement}. The output state $\rho'$ is analyzed using an MZI with two tunable phase shifters, allowing measurements in different bases. Finally, the signal and idler photons are coupled out of the chip via a fiber array and detected using superconducting nanowire single-photon detectors (SNSPDs). Notably, the silicon photonic chip is reconfigurable, allowing for experimental verification of ICO effects on quantum information transmission with $N=4$, $N=3$, $N=2$, and $N=1$ channels.

We experimentally verified the quantum information transmission capability for $N=1,2,3,4$. The four quantum states used as input states $\rho$ in our experiment are:$|D\rangle = \frac{1}{\sqrt{2}}(|0\rangle+|1\rangle)$, $|A\rangle = \frac{1}{\sqrt{2}}(|0\rangle-|1\rangle)$, $|R\rangle = \frac{1}{\sqrt{2}}(|0\rangle+i|1\rangle)$, $|L\rangle = \frac{1}{\sqrt{2}}(|0\rangle-i|1\rangle)$. 
We use \(|D\rangle, |A\rangle, |R\rangle, |L\rangle\) because they form two mutually unbiased bases that are sensitive to residual anisotropy or phase imbalance, and they coincide with the standard BB84 signal states, providing a quantum communication oriented probe of the channel.
The output states $\rho^{\prime}$ were reconstructed via quantum state tomography, and their fidelities were computed as follows
\begin{align}
\mathrm{F}_{\mathrm{q}}=\left( \mathrm{Tr}\left[ \sqrt{\sqrt{\rho}\cdot \rho ^{\prime}\cdot \sqrt{\rho}} \right] \right) ^2.
\end{align}
A detailed analysis of the control system's coherence and its influence on the fidelity measurements is provided in the Supplementary Material.
As shown in Fig.3, for $N=1$, where the quantum state passes through a single CDC, the measured fidelities for the four input states were $0.5003\pm0.0031$, $0.4995\pm0.0034$, $0.4992\pm0.0034$, and $0.5005\pm0.0034$, with error bars estimated by Monte Carlo simulations. These values correspond to a maximally mixed state, confirming that the channel cannot transmit any information.

For $N=2$ ($N=3$), i.e., when the quantum state propagates through two (three) CDCs via ICO, the measured fidelities were $0.599\pm0.0265$, $0.5964\pm0.0271$, $0.6014\pm0.0242$, and $0.5972\pm0.024$ for $N=2$, and $0.6632\pm0.0284$, $0.6629\pm0.0273$, $0.667\pm0.0245$, and $0.6665\pm0.0245$ for $N=3$. In both cases, the fidelities exceed 50\%, indicating that the channel can transmit classical information. However, since the fidelity does not surpass the classical threshold 2/3, the channel is still unable to transmit quantum information.

When $N=4$, i.e., when the quantum state propagates through four CDCs via ICO, the measured fidelities were $0.714\pm0.0292$, $0.7067\pm0.0256$, $0.7159\pm0.0238$, and $0.7095\pm0.0234$, which are in close agreement with the theoretical expectation of 0.7143. At this point, the fidelity significantly exceeds the classical threshold 2/3, demonstrating that the channel is now capable of supporting quantum information transmission.
 
\begin{figure}[tbp]
		  \centering\includegraphics[width=7.6cm]{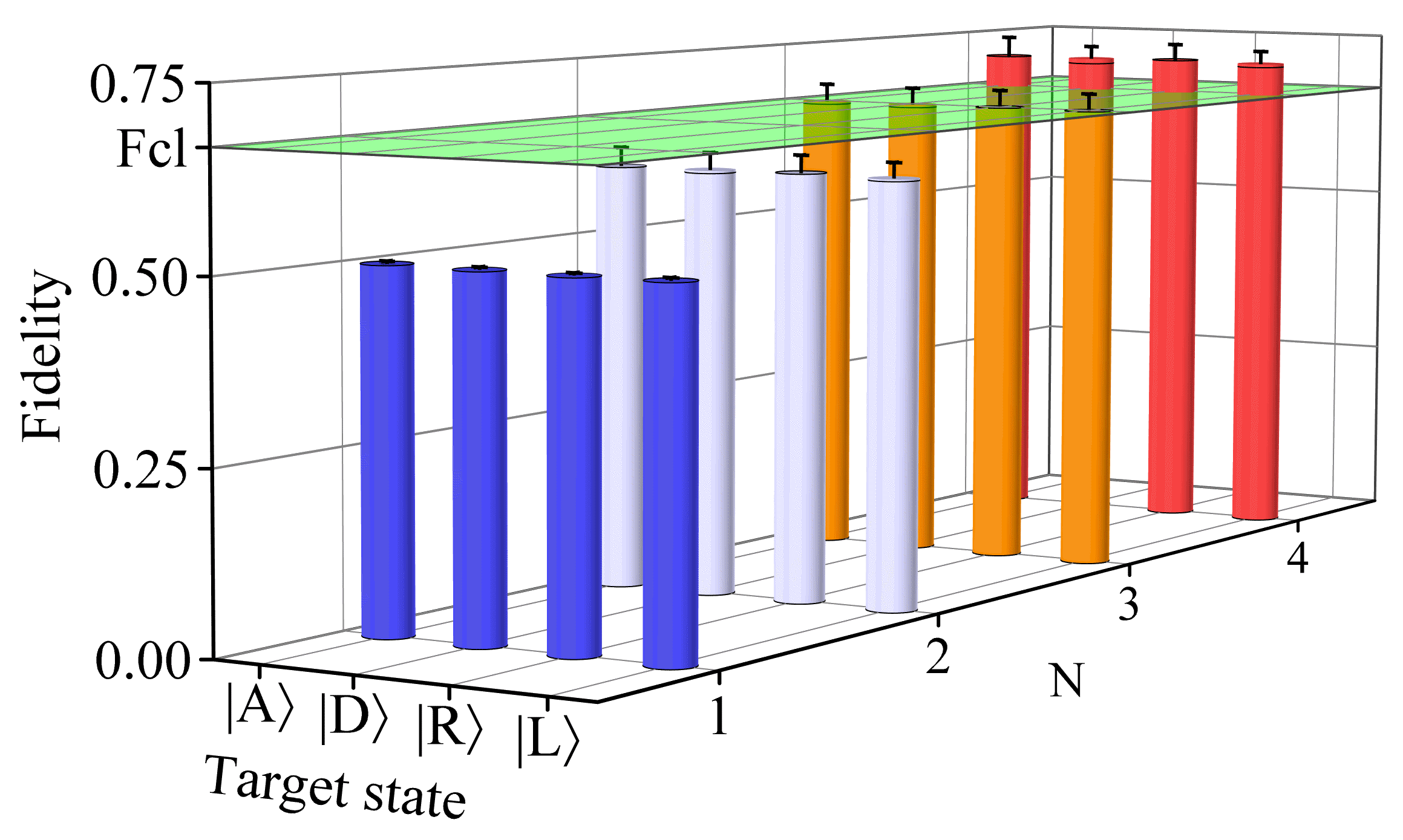}
		\caption{
      Experimental results of quantum state fidelity after propagation through multiple completely depolarizing channels under indefinite causal order. Four quantum states, $|D\rangle$, $|A\rangle$, $|R\rangle$, and $|L\rangle$, pass through $N=1,2,3,4$ completely depolarizing channels. When $N=4$, the average fidelity of the output states reaches $0.712 \pm 0.013$, surpassing the classical threshold of $2/3$.
        }
        \label{Fig3a}
	\end{figure}
\textbf{Discussion and conclusion.}---In summary, we report the first experimental realization of quantum information transmission through completely depolarizing channels (CDCs) via indefinite causal order (ICO). Using a programmable silicon photonic platform, we implemented a {superposition of cyclic order} structures to investigate its ability to restore quantum information in CDCs. Our results show that when a quantum state undergoes ICO through four CDCs, the output fidelity reaches $0.712 \pm 0.013$, exceeding the classical teleportation threshold 2/3, thereby confirming that the channel can support quantum information transmission. Beyond CDCs, the ICO-based approach may extend to other noise channels, such as phase-damping and non-Markovian channels. This technique has potential applications in quantum communication and distributed quantum computing, offering a robust framework to enhance quantum information processing.

\begin{acknowledgments}
This work was supported by the National Key research and development Program of China (No.2021YFA1400800), the National Natural Science Foundation of China (Grant No.62575313). X.-Q. Zhou acknowledges support from the Innovation Program for Quantum Science and Technology (Grant No.2021ZD0300702).
\end{acknowledgments}

\begingroup
\makeatletter\def\bibsection{}\makeatother 
\endgroup

\end{CJK*}
\end{document}